\documentclass[sigconf,screen, authorversion, nonacm]{acmart}

\usepackage{enumitem}
\usepackage{cleveref}

\setlist[itemize]{align=parleft,left=0.5em..1.5em}

\newlist{req}{enumerate}{2}
\setlist[req,1]{label=RQ \arabic*:,ref= \textbf{\arabic*}, leftmargin=*}

\AtBeginDocument{%
  }

\setcopyright{none}

\begin{document}

%\addbibresource{references}
%%
%% The ``title`` command has an optional parameter,
%% allowing the author to define a ``short title`` to be used in page headers.
\title{From Consumption to Collaboration: Measuring Interaction Patterns to Augment Human Cognition in Open-Ended Tasks}

\author{Joshua Holstein}
\email{joshua.holstein@kit.edu}
\affiliation{%
  \institution{Karlsruhe Institute of Technology}
  \city{Karlsruhe}
  \country{Germany}}

\author{Moritz Diener}
\email{moritz.diener@kit.edu}
\affiliation{
  \institution{Karlsruhe Institute of Technology}
  \city{Karlsruhe}
  \country{Germany}}

\author{Philipp Spitzer}
\email{philipp.spitzer@kit.edu}
\affiliation{
  \institution{Karlsruhe Institute of Technology}
  \city{Karlsruhe}
  \country{Germany}}

\renewcommand{\shortauthors}{Holstein et al.}

\begin{abstract}
The rise of Generative AI, and Large Language Models (LLMs) in particular, is fundamentally changing cognitive processes in knowledge work, raising critical questions about their impact on human reasoning and problem-solving capabilities. As these AI systems become increasingly integrated into workflows, they offer unprecedented opportunities for augmenting human thinking while simultaneously risking cognitive erosion through passive consumption of generated answers. This tension is particularly pronounced in open-ended tasks, where effective solutions require deep contextualization and integration of domain knowledge. Unlike structured tasks with established metrics, measuring the quality of human-LLM interaction in such open-ended tasks poses significant challenges due to the absence of ground truth and the iterative nature of solution development. To address this, we present a framework that analyzes interaction patterns along two dimensions: cognitive activity mode (exploration vs. exploitation) and cognitive engagement mode (constructive vs. detrimental). This framework provides systematic measurements to evaluate when LLMs are effective tools for thought rather than substitutes for human cognition, advancing theoretical understanding and practical guidance for developing AI systems that protect and augment human cognitive capabilities.
\end{abstract}




\maketitle

\section{Introduction}

The emergence of Generative AI (GenAI) has fundamentally reshaped how humans approach cognitive tasks, from creative ideation to complex problem-solving \cite{radensky2024scideator, Tankelevitch2024, simkute2025new}. Among these technologies, Large Language Models (LLMs) have become particularly influential due to their ability to engage in natural language dialogue and support a wide range of tasks \cite{subramonyam2024bridging}. While LLMs show remarkable potential for augmenting human thinking and reflection \cite{simkute2025new}, they simultaneously risk cognitive erosion through uncritical reliance on their outputs \cite{lee2025impact, sarkar2024copilot, drosos2025makes, stadler2024cognitive}. This risk is particularly pronounced as LLMs can generate seemingly high-quality outputs that may contain substantial flaws \cite{russell2024sensemaking}: they can produce hallucinations---confidently stated but factually incorrect information---that can lead to misinformation and flawed mental models \cite{spitzer2024don}. The fluent articulation of these potentially problematic outputs makes them especially susceptible to uncritical acceptance by humans \cite{Tankelevitch2024}.

Current research proposes various design approaches to prevent cognitive erosion and promote meaningful engagement with LLMs. These approaches aim to facilitate reflection and critical thinking through different mechanisms, including specialized RAG-systems \cite{radensky2024scideator}, learning environments \cite{chen2024learning}, or hierarchical recommendations with prompt templates \cite{zhang2023visar}. These design interventions are particularly relevant in open-ended tasks---such as strategic business choices, career development planning, and complex design decisions---where ambiguous goals and multiple solution paths require deep cognitive engagement to develop well-contextualized solutions \cite{lai2011critical}. Yet, unlike tasks with well-defined evaluation criteria, such as mathematical problem-solving \cite{collins2024evaluating} or code generation \cite{Vasconcelos2024, chen2024learning}, measuring the success in these open-ended tasks poses significant challenges.

While prior research has extensively studied the success of human-AI collaboration in traditional AI systems for classification \cite{Schemmer2023, schoeffer2024explanations, lu2021human, wang2023watch, morrison2024impact} and regression tasks \cite{hemmer2022effect, kahr2024understanding}, initial work has started to investigate human-LLM collaboration in structured domains like code generation or math \cite{Vasconcelos2024, collins2024evaluating}. However, in contrast to ``traditional'' AI systems where an objective ground truth exists, dialogues between humans and LLMs in open-ended tasks create complex interaction patterns---such as iterative refinement of ideas, exploration of alternative solutions, and critical evaluation of LLM suggestions---that are inherently more difficult to evaluate \cite{Tankelevitch2024}. These patterns emerge as humans engage with LLMs through multiple rounds of questioning, refinement, and synthesis, often involving both creative exploration and detailed elaboration of specific approaches. Yet, without an understanding of these interaction patterns, we lack the ability to design and validate LLM systems that genuinely enhance human cognitive processes rather than diminish them. This gap leads to our research question:

\begin{req}[labelindent=1em, labelwidth=0em, label=\textbf{RQ}:, ref=\arabic*]
    \item \textit{How can we conceptualize and measure interaction patterns in human-LLM collaboration to distinguish cognitive augmentation from erosion in open-ended tasks?}
    \label{rq1}
\end{req}

To answer our research question, we introduce a framework that conceptualizes human-AI collaboration patterns along two fundamental dimensions: \textbf{cognitive activity mode} (exploration vs. exploitation) and \textbf{cognitive engagement mode} (constructive vs. detrimental). By distinguishing between these dimensions, our framework provides a structured approach to understanding when LLMs effectively augment human cognition and when they risk passive engagement. This work contributes to both theoretical discourse and practical system design, offering measurable criteria for evaluating and promoting meaningful cognitive engagement in open-ended tasks.

\section{Conceptualization}
To analyze how humans engage with LLMs in open-ended tasks, we develop a framework building on established theories of exploration-exploitation trade-off \cite{march1991exploration} and cognitive engagement \cite{greene2015measuring}.

\subsection{Cognitive Activity Mode: The Exploration-Exploitation Trade-off}
Cognitive activity modes represent a fundamental tension in human learning and decision-making, characterized by the trade-off between exploration and exploitation \cite{march1991exploration}. This trade-off manifests as a continuous process of balancing exploring new possibilities with exploiting existing solutions. In open-ended tasks, where optimal solutions are unclear and context-dependent, this balance becomes particularly critical: Exploration enables discovering diverse solution paths and challenging assumptions, while exploitation allows deep development of promising approaches through iterative refinement.

\textit{\textbf{Exploration} describes a cognitive process of investigating unknown solutions, characterized by experimentation and discovery under uncertainty \cite{march1991exploration, harada2023exploring}.} In the context of human-LLM interaction for open-ended tasks, exploration manifests in activities where humans actively seek new perspectives, challenge assumptions, and generate novel ideas. Humans typically engage in divergent thinking \cite{harada2023exploring}, using the LLM as a tool for generating novel ideas. This might involve probing the system with open-ended questions, deliberately seeking alternative viewpoints, or using the LLM's responses as prompts for further creative thinking. Exploration is particularly valuable in the early stages of open-ended tasks, where the problem space needs to be thoroughly understood and various solution paths considered.

\textit{\textbf{Exploitation} represents a cognitive process focused on utilizing and refining existing knowledge, characterized by the systematic improvement and application of identified solutions \cite{march1991exploration, harada2023exploring}.} In the context of human-LLM interaction for open-ended tasks, exploitation involves leveraging the system's capabilities to refine and implement existing ideas or solutions. Humans in this mode typically engage with the LLM to deepen their understanding of specific solutions, seek detailed implementation guidance, or validate and improve already identified approaches. This might include requesting clarification of complex concepts, exploring the implications of chosen solutions, or refining specific aspects of an approach. Exploitation becomes particularly important when promising directions are identified and must be further developed to create practical, implementable solutions.

These two cognitive activities---exploration and exploitation---form a spectrum along which humans navigate when engaging with LLMs for problem-solving and the balance between them shapes the overall quality and depth of human-LLM collaboration. While exploration demands higher cognitive effort through active engagement with novel solutions, exploitation offers a potentially less demanding path of iteratively refining existing ones. When interacting with LLMs, this dynamics become particularly critical---the system's ability to generate seemingly good and coherent responses might unconsciously incline humans toward exploitation \cite{russell2024sensemaking}, potentially limiting both their cognitive engagement and the depth of the developed solution.

\subsection{Cognitive Engagement Mode: The Constructive-Detrimental Quality}

The mode of cognitive engagement \cite{greene2015measuring} represents our second dimension in understanding how humans interact with LLMs, particularly in open-ended tasks where deep thinking and critical reflection are essential to developing well-contextualized solutions.

\textit{\textbf{Constructive Engagement} describes cognitive processes where humans invest cognitive effort to maintain active agency and critical thinking \cite{greene2015measuring}.} In open-ended tasks, constructive engagement manifests when humans thoughtfully evaluate LLM outputs, integrate system suggestions with their own knowledge and experience, and strategically use the LLM to extend their thinking capabilities. For instance, when an LLM suggests a solution approach, constructively engaged humans actively contextualize this suggestion within their specific problem context, critically examining its assumptions and limitations. They might follow up with probing questions to understand the reasoning behind suggestions, deliberately challenge the system's responses, or use them as starting points for developing more detailed solutions that incorporate their domain expertise. This type of engagement ensures that the LLM serves as a thought partner rather than a replacement for human judgment.

\textit{\textbf{Detrimental Engagement} represents patterns of shallow, mechanical interactions where humans' cognitive processes are diminished rather than enhanced \cite{greene2015measuring}.} In open-ended tasks, this typically occurs when humans uncritically accept LLM outputs, overrely on the system for cognitive tasks, or engage in passive consumption of information without processing or integrating them into their knowledge and experience or tasks at hand. This form of engagement is particularly concerning as it may create an illusion of productivity while actually undermining the development of robust problem-solving capabilities and domain understanding, potentially even leading to a decrease in the quality of generated outputs.

\subsection{Framework Integration: A Two-Dimensional View of Human-LLM Interaction}

\begin{figure}[h!]
    \centering
    \includegraphics[width=0.75\linewidth]{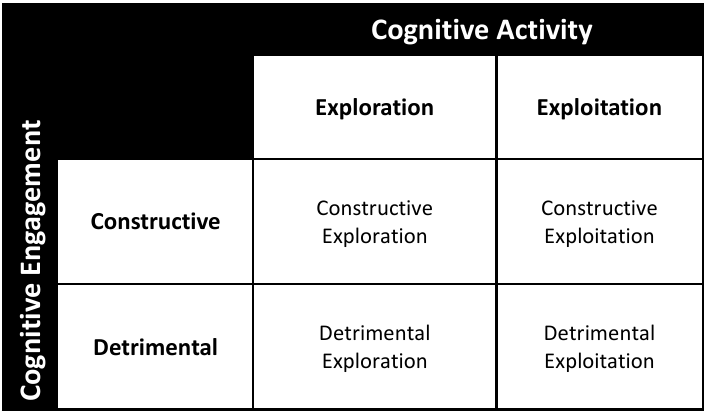}
    \caption{We distinguish between four types of human-LLM interaction across the two dimensions of cognitive activity mode and cognitive engagement mode.}
    \label{fig:framework}
\end{figure}

By integrating the cognitive mode with the quality of cognitive engagement, we develop a comprehensive framework for analyzing human-LLM interaction patterns in open-ended tasks (see \Cref{fig:framework}). This integration creates four distinct interaction patterns, each with unique implications for cognitive development and task outcomes.

\begin{itemize}
    \item \textbf{Constructive Exploration} characterizes interactions where humans maintain high cognitive engagement while exploring new possibilities. Humans leverage the LLM to enhance creative thinking and problem-solving while maintaining critical judgment and synthesizing multiple perspectives with their own knowledge. This pattern is particularly valuable when the problem space needs a thorough understanding or existing approaches prove insufficient.

    \item \textbf{Constructive Exploitation} occurs when humans maintain thoughtful engagement while refining specific solutions. humans leverage the LLM to deepen their understanding while actively contextualizing and evaluating suggestions, engaging in detailed dialogue about specific aspects while ensuring alignment with their broader goals. This pattern becomes valuable when promising directions require further careful contextualization and development.
    
    \item \textbf{Detrimental Exploration} manifests as superficial brainstorming or unfocused idea generation without deep processing or critical evaluation. humans might accumulate numerous suggestions from the LLM without meaningfully engaging with any of them, leading to scattered thinking and shallow understanding. This pattern creates an illusion of productive exploration while failing to develop genuine insights or viable solutions.
    
    \item \textbf{Detrimental Exploitation} represents the most concerning pattern, characterized by excessive reliance on the LLM and minimal cognitive contribution from the user. In this mode, humans might uncritically implement the LLM's suggestions, delegate core thinking tasks to the system, or focus solely on refining LLM-generated solutions without meaningful evaluation or integration with their knowledge. This pattern risks producing suboptimal solutions and eroding humans' cognitive capabilities over time.
\end{itemize}

%This framework provides a structured approach for analyzing and promoting effective human-LLM collaboration in open-ended tasks. By understanding these interaction patterns, we aim to design better LLM systems and interfaces that encourage constructive engagement while mitigating the risks of cognitive erosion.
Through understanding these interaction patterns, we aim to design LLM systems that foster constructive engagement while preventing cognitive erosion

\section{Framework Application}
To move from theoretical conceptualization to practical application, we develop a systematic approach for analyzing human-LLM interactions in open-ended tasks. This section presents our measurement methodology and demonstrates its application through an illustrative example.

\subsection{Measurement Approach}
Building on our conceptualization of cognitive activity and cognitive engagement mode, we systematically measure human-LLM interaction patterns in open-ended tasks.

\textbf{Qualitative Analysis of Cognitive Activity Mode.} We first segment dialogues based on transitions between exploration and exploitation modes. A new segment begins when humans switch their cognitive mode, for instance, from exploring alternative solutions (e.g., seeking unconsidered perspectives) to exploiting a specific approach (e.g., detailed elaboration or implementation guidance). This segmentation enables us to analyze both the frequency and duration of different cognitive modes throughout the dialogue. By calculating the ratio of exploration segments to total segments, we derive a score between 0 and 1 that quantifies the balance between exploration and exploitation, where 0 represents pure exploitation (all segments focused on refining specific solutions) and 1 represents pure exploration (all segments dedicated to discovering new possibilities). This measurement approach allows us to evaluate how different interface designs or interventions might influence the sequence and duration of exploration and exploitation phases.

\textbf{Analysis of Cognitive Engagement Mode.} Within each segment, we assess cognitive engagement mode through three key indicators: (1) the presence and quality of follow-up questions that build on previous responses (e.g., probing for deeper explanations or requesting clarification of assumptions), (2) explicit integration of personal context or experience (e.g., relating LLM suggestions to specific situations or domain knowledge), and (3) critical examination of LLM suggestions (e.g., challenging proposed solutions or identifying potential limitations). These indicators help classify each segment as either constructive or detrimental engagement. Constructive engagement is characterized by active processing and thoughtful integration of LLM outputs, while detrimental engagement manifests as passive consumption without meaningful evaluation or contextualization. Similar to our cognitive activity mode analysis, we calculate the ratio of constructively engaged segments to total segments, resulting in a score between 0 and 1, where 0 represents purely detrimental engagement and 1 represents consistently constructive engagement throughout the dialogue. This measurement approach enables us to identify patterns of cognitive engagement and evaluate how different interaction designs might promote more constructive engagement patterns.

\textbf{Automated Analysis Through LLMs.} To enable systematic analysis of larger datasets, we propose to leverage LLMs as analytical tools \cite{feuerriegel2025using}. First, we prompt LLMs to identify segment transitions by recognizing shifts between exploration and exploitation modes. Second, we prompt them to assess cognitive engagement mode within each segment. This automation enables consistent analysis across diverse interaction logs while reducing the resource intensity of manual coding

\subsection{Demonstration and Analysis}

\begin{figure}
    \centering
    \includegraphics[width=1\linewidth]{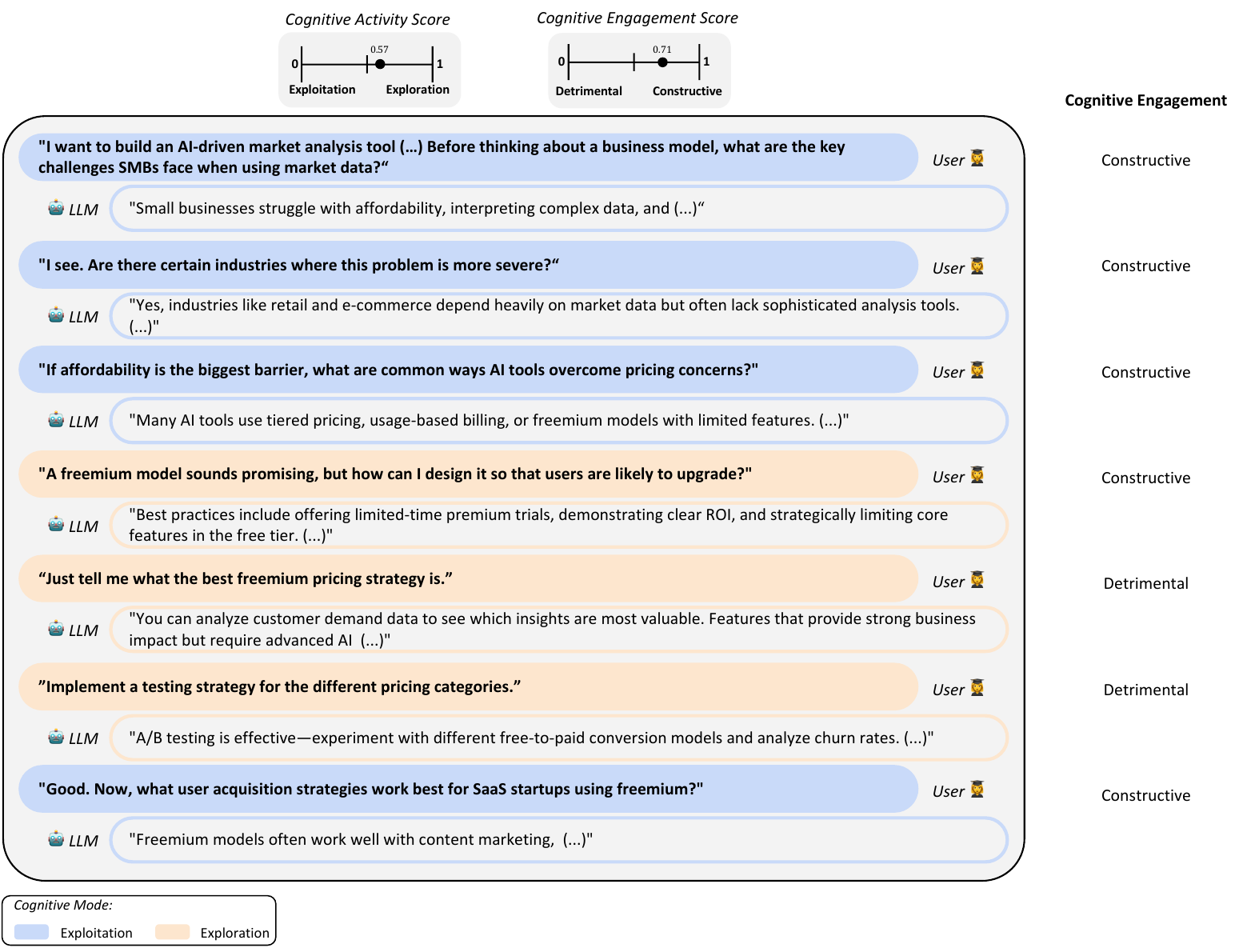}
    \caption{An exemplary chat history, where an LLM coded the user interaction along our two dimensions: cognitive activity and cognitive engagement mode.}
    \label{fig:example}
\end{figure}

To illustrate how our framework can be applied to analyze human-LLM interactions, consider a dialogue where a user explores building an AI-driven market analysis tool (see \Cref{fig:example}). The dialogue structure was designed to simulate authentic user interactions, incorporating both exploratory and exploitative approaches while demonstrating different patterns of engagement. To code the respective interactions, we prompted GPT-4o (version of November 20, 2024) to classify the interactions based on our definitions. 

The interaction begins with constructive exploration as the user seeks to understand fundamental challenges (``What are the key challenges SMBs face when using market data?'') and follows up with contextual questions about industry-specific issues. This exploration phase maintains a high cognitive engagement through strategic questioning and building upon previous responses. A transition to constructive exploitation occurs when the user narrows focus to pricing strategy (``A freemium model sounds promising, but how can I design it?''). Here, the user actively engages with a specific solution while maintaining critical thinking through targeted questions. However, the interaction quality deteriorates into detrimental exploitation when the user shifts to passive consumption (``Just tell me what the best freemium pricing strategy is'') and directive statements (``Implement a testing strategy''), showing minimal cognitive contribution. The dialogue concludes with a return to constructive exploration as the user broadens the scope to consider user acquisition strategies, demonstrating how interaction patterns can shift throughout a conversation. Using our measurement approach, we can quantify these patterns: the exploration-exploitation ratio indicates a balanced distribution with 4 out of 7 prompts being exploratory (57.14\% exploration), while the cognitive engagement mode score (71.43\% constructive) reveals predominantly meaningful cognitive engagement despite periods of passive consumption.

This example illustrates how our framework can identify and measure different interaction patterns, providing insights into when and how humans maintain constructive engagement versus falling into more passive interaction modes. The relatively balanced ex\-plo\-ra\-tion-exploitation ratio (57.14\%) suggests a healthy dynamic between discovering new possibilities and deepening specific solutions, while the high proportion of constructive engagement (71.43\%) indicates that the user maintained meaningful cognitive contribution throughout most of the dialogue. However, the observed transition into detrimental exploitation during the pricing strategy discussion highlights how easily interaction quality can deteriorate when users seek quick solutions rather than maintaining thoughtful engagement. These measurements enable systematic evaluation of interaction quality and could inform the design of interventions that help users maintain constructive engagement patterns, particularly during transitions between exploration and exploitation phases.

\section{Discussion}
Our framework contributes to the theoretical understanding of human-LLM interaction as generative AI becomes increasingly prevalent in knowledge work. By systematically distinguishing between cognitive activity modes and engagement patterns, we extend beyond traditional effectiveness metrics to examine the underlying mechanisms of human-LLM interaction. While LLMs demonstrate significant potential for augmenting human thinking \cite{simkute2025new}, our framework reveals how their cognitive impact fundamentally depends on the quality of user engagement: the same system can either enhance or diminish human cognitive capabilities based on whether it facilitates active reasoning or enables passive consumption \cite{stadler2024cognitive}. Through automated analysis, our framework enables real-time detection of engagement patterns and shifts toward detrimental interactions, opening avenues for timely interventions that ensure LLMs augment rather than diminish critical thinking.

The practical implications of our framework inform the design of AI systems that protect and augment human thinking. Our measurement approach enables systematic evaluation of how interface elements influence cognitive engagement, allowing for adaptive interventions when detrimental patterns are detected. Such interventions could prompt humans to articulate reasoning, consider alternative perspectives during exploration, or deepen reflection during exploitation phases. These design principles become increasingly important as AI systems integrate into professional and educational cognitive workflows.

Our framework's primary limitation is its need for empirical validation across different domains and contexts. Future studies should investigate how interface designs influence both cognitive activity modes and engagement patterns in open-ended tasks, providing evidence-based guidelines for LLM interfaces that enhance rather than diminish human cognitive capabilities. Finally, empirical validation must also verify the reliability and validity of our proposed measurement approach, particularly its ability to identify and distinguish between different interaction patterns reliably and whether these patterns translate in varying output quality.

\section{Conclusion}
Understanding how to design generative AI systems that genuinely augment human cognition while preventing cognitive erosion is a critical challenge of our time. Our framework offers a systematic approach to analyzing and measuring human-LLM interaction patterns in open-ended tasks, distinguishing between constructive and detrimental engagement forms across exploration and exploitation modes. By providing measures and design implications, this work contributes to the development of generative AI systems that enhance rather than diminish human cognitive capabilities. As LLMs become increasingly integrated into cognitive workflows, such frameworks become essential for ensuring these powerful tools augment human thinking rather than replace it.

\bibliographystyle{ACM-Reference-Format}
\bibliography{references}
\end{document}